# Capacity of Two-Way Relay Channel

Shengli Zhang, Soung Chang Liew
The Department of Information Engineering,
The Chinese University of Hong Kong, Hong Kong, China
Email: {slzhang5, soung}@ie.cuhk.edu.hk

*Abstract:* This paper investigates the capacity of a wireless two-way relay channel in which two end nodes exchange information via a relay node. The capacity is defined in the information-theoretic sense as the maximum information exchange rate between the two end nodes. We give an upper bound of the capacity by applying the cut-set theorem. We prove that this upper bound can be approached in low SNR region using "separated" multiple access for uplinks from the end nodes to the relay in which the data from the end nodes are individually decoded at the relay; and network-coding broadcast for downlinks from the relay to the end nodes in which the relay mixes the information from end nodes before forwarding. We further prove that the capacity is approachable in high SNR region using physical-layer network coding (PNC) multiple access for uplinks, and network-coding broadcast for downlinks. From our proof and observations, we conjecture that the upper bound may be achieved with PNC in all SNR regions.

## I. INTRODUCTION:

The design and analysis of wireless two-way relay channel (TWRC) are attracting increasing attention. Among all the models for TWRC, of greatest interest is the simplest model of three-node TWRC with additive Gaussian noise. This paper focuses on a three-node TWRC system as depicted in Fig. 1. We investigate its maximum information exchange rate, defined as the smaller of the forward-direction and reverse-direction information transfer rates.

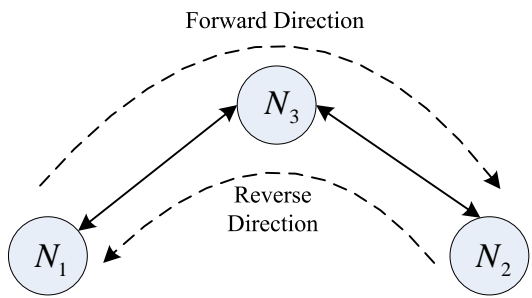

Figure 1. Three-node two way relay channel

The two-way channel without a relay was first studied by Shannon in [1]. The one-way relay channel under various relay strategies has also been widely studied (see [2] and the references therein). As a combination of the two, TWRC is currently drawing much research attention. Ref. [3] gave the achievable rate regions of TWRC when relay protocols borrowed from the one-way relay channel are applied. Under the same relay protocols, [4] derived the rate regions of the TWRC when the nodes are half-duplex and a direct link between the two ends is not available. These strategies that do not exploit network coding cannot achieve the information transfer rate at capacity. Application of network coding is not very fruitful for one-way relay channel. However, applying network coding in TWRC opens up new possibilities.

As shown in [5], broadcasting a network-coded version of the inputs at the relay node holds the promise of significant system-performance improvement. Based on this idea, practical joint network coding and channel coding designs were studied in [6, 7]. Ref. [8] proposed a "physical layer network coding" (PNC) scheme which further improves the transmission efficiency of TWRC. Other transmission schemes inspired by PNC were proposed and analyzed in [9, 10, 11] for TWRC. To our knowledge, all the current papers on TWRC focus on the achievable rate regions or bounds under specific transmission strategies. The "ultimate" achievable maximum rate region of TWRC is still an open issue.

In contrast to previous work, this paper studies the capacity of TWRC as the maximum information exchange rate under all the possible transmission strategies. In particular, we give an upper bound on symmetric information exchange rate based on the cut-set theorem. We prove that this upper bound is approachable in low SNR region using "separated" multiple-access for uplinks and network-coding broadcast for downlinks. In separated multiple-access [13], the relay node extracts the "complete" information from each of the two nodes; the "separated" information is then combined using network coding [5] for downlink broadcast. In high SNR region, the upper bound is approachable by replacing separated multiple access with PNC multiple access. We conjecture that the upper bound is achievable in all SNR regions with PNC.

The paper is organized as follows. Section II introduces the system model studied in this paper. Section III gives upper bound of the system capacity. Section IV and V prove the tightness of the upper bound in low SNR and high SNR regions, respectively. Section VI concludes this paper.

## II. SYSTEM MODEL:

With reference to Fig. 1, nodes $N_1$ and $N_2$ exchange information with the help of node $N_3$. We assume all nodes are half-duplex, i.e., each node can not receive and transmit simultaneously. This is an assumption arising from practical considerations because it is difficult for wireless nodes to remove the strong interference of its own transmitting signal from the received signal. We also assume that there is no direct link between nodes $N_1$ and $N_2$. A practical example is satellite communication, wherein two end nodes on the earth can only communicate with each other via a satellite relay.

We assume that the maximum transmission power of node $N_i$ is $P_i$, and receiver noise is addictive Gaussian with unit variance at all nodes. The path loss is unit constant. By low SNR region, we mean both $P_1$ and $P_2$ approach zero, and by high SNR region we mean both $P_1$ and $P_2$ approach infinity. We do not consider fading in this paper.

*Notations*

In this paper, $n$ denotes the Gaussian noise. $W_i$ denote the information packet of node $N_i, i \in \{1,2,3\}$, and $w_i[k] \in \{0,1,\cdots q-1\}, k \in \{0,1\cdots K-1\}$ denote its $k$-th symbol. The $q$-ary information symbols are used throughout the paper.

For channel coding, $\Gamma_1$, $\Gamma_2$ and $\Gamma_3$ denote the encoding functions of $N_1$, $N_2$, and $N_3$ respectively. The coded packets of node $i$ is

$$U_i = \Gamma_i(W_i) \quad , \quad i \in \{1,2,3\}$$

Within $U_i$, $u_i[k] \in \{0,1,\cdots q-1\}, k \in \{0,1\cdots L_i-1\}$ denote its $k$-th symbol. The coding rate of $N_i$ is therefore $K/L_i$.

We assume all nodes use the same modulation scheme. $X_i$ denote the modulated packet of $N_i$, and $M$ denote the modulation mapping:

$$X_i = M(U_i) \quad i \in \{1,2,3\}$$

Within $X_i$, $x_i[k] \in \{0,1,\cdots q-1\}$, $k \in \{0,1\cdots L_i-1\}$ denote its $k$-th modulated symbol. Finally, $y_i[k]$ denote the $k$-th received baseband signal at node $N_i$.

The capacity of the system is defined as

$$C = \max_{s \in \{all\ possible\ schemes\}} \min\{R_{1,2}(s), R_{2,1}(s)\} \quad (1)$$

where $R_{2,1}(s)$ is the reliable transmission rate from $N_2$ to $N_1$ under transmission scheme $s$, and $R_{1,2}(s)$ is the transmission rate in the opposite direction during the same time under the same transmission scheme.

### III. UPPER BOUND OF THE TWRC CAPACITY:

The upper bound of the TWRC capacity defined in (1) is given in the following proposition with a cut-set proof.

*Proposition 1:* For the TWRC model described in the previous section, the capacity defined in (1) is upper-bounded by

$$C \leq \frac{1}{2} \frac{\log_2(1+\min(P_1,P_2)) \cdot \log_2(1+P_3)}{\log_2(1+\min(P_1,P_2)) + \log_2(1+P_3)} \quad (2)$$

where $\frac{1}{2}\log_2(1+P_i)$ is the Shannon channel capacity for a Gaussian channel with SNR $P_i$.

**Proof:** Due to half-duplexity and the lack of a direct link between $N_1$ and $N_2$, it is necessary to divide the transmission into two phases, one phase for $N_3$'s reception and the other phase for $N_3$'s transmission. The first phase is referred to as the uplink phase, which includes all the transmissions from node $N_1$ and/or $N_2$ to node $N_3$. Note that there are at most three possible transmission scenarios: $N_1$ transmits to $N_3$; $N_2$ transmits to $N_3$; or $N_1$ and $N_2$ transmit to $N_3$ simultaneously. In this phase, the maximum reliable transmission rate originating from $N_1$ is no more than $\frac{1}{2}\log_2(1+P_1)$. This result is obtained by applying the cut-set theorem where $N_1$ is regarded as the source set as in [20]. Similarly, the maximum reliable transmission rate from $N_2$ is no more than $\frac{1}{2}\log_2(1+P_2)$. Since we are interested in the smaller of the two rates, the transmission rate in the uplink phase is upper bounded by $\frac{1}{2}\log_2(1+\min(P_1,P_2))$. The second phase is referred as the downlink phase, which includes all the transmissions originating from $N_3$. According to the Shannon channel capacity, the information transmission rate from $N_3$ to $N_1$ and/or $N_2$ is no more than $\frac{1}{2}\log_2(1+P_3)$. Besides the transmission rates of the two phases, the final exchange rate also depends on the time allocation. We use $t_1$ to denote the total time used in the first phase, and $1-t_1$ to denote the total time used in the second phase. Then the final information exchange rate is

$$\min\left\{\frac{t_1}{2}\log_2(1+\min(P_1,P_2)), \frac{1-t_1}{2}\log_2(1+P_3)\right\} \quad (3)$$

To maximize the value in (3), obviously we should determine the value of $t_1$ so that the two arguments in the min function are equal. Doing so yields the upper bound in (2).

### IV. APPROACHING THE UPPER BOUND IN LOW SNR REGION

With reference to the proof of *Proposition 1*, we must find a transmission scheme that can approach the upper bounds in both phases to prove the tightness of the upper bound in (2). Let us first consider the downlink phase.

*Downlink Phase*

Our scheme consists of two steps. The first step is to decide a function $W_3=f(W_1, W_2)$. After that, we could use a standard capacity approaching channel coding scheme, such as the LDPC code or the Turbo code, to encode the information $W_3=f(W_1, W_2)$ into $U_3$ before modulating and broadcasting it to both $N_1$ and $N_2$ with a rate approaching $\frac{1}{2}\log_2(1+P_3)$. Fig.2 shows this transmission strategy from the viewpoint of reception at $N_1$.

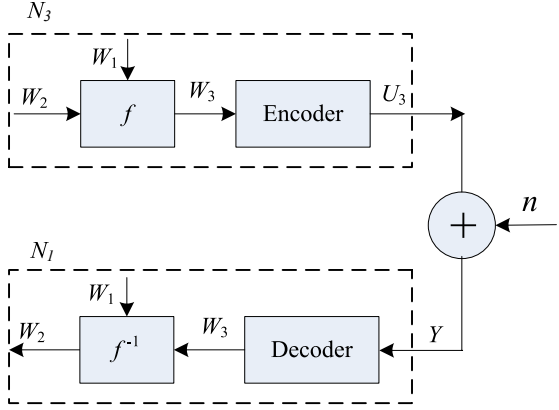

Figure 2. Reception of downlink transmission at node $N_1$

Since we use a capacity approaching channel code, $N_1$ and $N_2$ can obtain $W_3$ with an error probability approaching zero as long as the rate of $W_3$ is no more than $\frac{1}{2}\log_2(1+P_3)$. The critical issue for the downlink transmission is to find a function $f$ such that the target information $W_1$ and $W_2$ can be decoded from $W_3$ at both end nodes with the same rate of $W_3$. The requirement can be specified as follows:

***Requirement to Satisfy the Downlink Phase Upper Bound:***

$$H(W_2) = H(W_1) = H(W_3)$$

where $W_1$ and $W_2$ are decodable from $W_3$ at $N_2$ and $N_1$ respectively.

With reference to the receiving part in Fig. 2, the interpretation of this requirement is as follows. In order that the transmission rate of $W_2$ can achieve the upper bound, there should be no information loss during the signal processing to obtain $W_2$ from $W_3$. In other words, the entropy of $W_2$ obtained at $N_1$ should be equal to that of $W_3$. To meet this requirement, the following proposition presents the necessary and sufficient conditions that $f$ must satisfy.

***Proposition 2:*** The function $f$ must satisfy both of the following conditions to achieve the upper bound of the downlink phase with the given downlink transmission strategy.

$$H(W_2 \mid W_1, W_3) = 0 \quad H(W_1 \mid W_2, W_3) = 0 \quad (4)$$

$$I(W_3; W_1) = 0 \quad I(W_3; W_2) = 0 \quad (5)$$

**Proof:** With reference to the receiving part in Fig. 2, we have the following inequalities according to basic Information Theory rules,

$$H(W_2) = H(W_2 \mid W_1) \leq H(W_3 \mid W_1) \leq H(W_3) \quad (6)$$

The function $f$ must satisfy the inequalities in (6) with equality. The satisfaction of the first inequality $H(W_2 \mid W_1) \leq H(W_3 \mid W_1)$ at equality is equivalent to condition (4), as shown below:

$$\begin{aligned}
&H(W_2 \mid W_1) - H(W_3 \mid W_1) \\
&= H(W_2 \mid W_1) - H(W_3 \mid W_1) + H(W_3 \mid W_1, W_2) \\
&= H(W_2 \mid W_1) - I(W_3; W_2 \mid W_1) \\
&= H(W_2 \mid W_1) - H(W_2 \mid W_1) + H(W_2 \mid W_3, W_1) \\
&= H(W_2 \mid W_3, W_1)
\end{aligned} \quad (7)$$

The satisfaction of the second inequality $H(W_3 \mid W_1) \leq H(W_3)$ at equality is equivalent condition (5):

$$H(W_3) - H(W_3 \mid W_1) = I(W_3; W_1) \quad (8)$$

When $N_2$ is considered, we can obtain similar results.

Condition (4) means that $f$ must be reversible when $W_1$ or $W_2$ is given. Otherwise, the end nodes cannot recover their counterpart's information from $W_3$ and their self information. Condition (5) means that the output of $f$ must be probabilistically independent of each of the two input packets alone. Otherwise, the downlink transmission in Fig. 2 is inefficient. Both conditions are needed. For example, the function $f(W_1, W_2) = W_1 + W_2$ satisfies (4) but not (5). On the other hand, $f(W_1, W_2) = W_0$, where $W_0$ is a random packet, satisfies (5) but not (4). Neither of these functions can achieve the upper bound of the downlink phase. For an $f$ that satisfies both (4) and (5), consider the network coding operation over a finite field, $f(W_1, W_2) = W_1 \oplus W_2$, as in [12].

In summary, in the scheme of Fig. 2 in which the channel coding and signal processing of $f$ are performed separately, the upper bound of the downlink transmission rate can be achieved with a valid $f$ and a capacity-approaching channel code. We refer to such a downlink scheme as network-coding broadcasting. In the remainder of this paper, $f$ denotes a valid function that can achieve the upper bound in the downlink phase.

*Uplink Phase*

We now consider the uplink phase. We need to find a multiple-access scheme with which we can obtain $f(W_1, W_2)$ with a rate approaching $\frac{1}{2}\log_2(1+\min(P_1, P_2))$ in the uplink phase. In fact, the separated multiple-access scheme in [13] guarantees that $N_3$ can obtain both $W_1$ and $W_2$ with a rate at least approaching $\frac{1}{2}\log_2(1+\min(P_1, P_2))$ in low SNR region, from which $W_3 = f(W_1, W_2)$ can then be computed.

***Proposition 3:*** In low SNR region, $N_3$ can obtain both $W_1$ and

$W_2$ at a rate approaching $\frac{1}{2}\log_2(1+\min(P_1,P_2))$ with the help of the separated multiple-access scheme.

**Proof:** Without loss of generality, we assume that $L_1 \geq L_2$, which implies $P_1 \leq P_2$. The two end nodes cooperate with each other to transmit at the same time to $N_3$. Then $N_3$ decodes $W_2$ from the received superimposed packet $Y_3$. Treating $N_1$'s information $W_1$ as interference, $N_3$ can reliably decode $W_2$ with a rate approaching $\frac{1}{2}\log_2(1+P_2/(P_1+1))$. After that, $N_3$ can decode $W_1$ after removing the information of $W_2$ from $Y_3$. As a result, the reliable transmission rate of $W_1$ approaches $\frac{1}{2}\log_2(1+P_1)$.

If $P_2/(P_1+1) \geq P_1$, i.e., $P_2 \geq P_1 + P_1^2$, the rate of $W_1$, which is smaller than that of $W_2$, approaches $\frac{1}{2}\log_2(1+\min(P_1,P_2))$. On the other hand, if $P_1 \leq P_2 \leq P_1 + P_1^2$, the rate of $W_2$, which is smaller than that of $W_1$, can be approximated by

$$\frac{1}{2}\log_2(1+P_2/(P_1+1))$$
$$=\frac{1}{2}\log_2(1+P_1+\frac{P_2-P_1-P_1^2}{P_1+1}) \quad (9)$$
$$\geq \frac{1}{2}\log_2(1+P_1-\frac{P_1^2}{P_1+1}) \xrightarrow{P_1 \to 0} \frac{1}{2}\log_2(1+P_1)$$

Hence, *Proposition 3* is proved.

Since we can obtain both $W_1$ and $W_2$ at a rate of $\frac{1}{2}\log_2(1+\min(P_1,P_2))$, we can obtain $f(W_1, W_2)$ (e.g., the summation of $W_1$ and $W_2$ over a finite field) at the same rate. With the help of separated multiple access and network-coding broadcast, we can approach the upper bound of the TWRC' capacity in (2) in low SNR region. This result indicates that the traditional network coding, which regards network coding as an upper layer operation (i.e., $W_1$ and $W_2$ are decoded explicitly at $N_3$ before $f(W_1, W_2)$ is computed from $W_1$ and $W_2$ at the upper layer) is near optimal in low SNR region in TWRC. However, this is not the case in high SNR region.

## V. APPROACHING THE UPPER BOUND IN HIGH SNR REGION:

As for low SNR region, we also need to find a multiple-access scheme that allows $N_3$ to obtain $f(W_1, W_2)$ with a rate approaching $\frac{1}{2}\log_2(1+\min(P_1,P_2))$ in the uplink phase to prove the tightness of the upper bound in (2). We find that tradition separated multiple access does not work any more. In this section, we show that PNC-based multiple access can approach the upper bound.

For PNC multiple access, we use an *f* that is fixed during the whole downlink phase, as follows:

$$f(W_1,W_2) = W_1 +_q W_2 = W_1 + W_2 \bmod q \quad (10)$$

where $+_q$ denotes the modulo-$q$ addition (Modulo-$q$ addition of two packets means symbol-wise modulo-$q$ addition). It is easy to verify that the function in (10) satisfies the conditions (4) and (5). Thus, the downlink phase can approach the rate of $\frac{1}{2}\log_2(1+P_3)$. The following proposition shows that for the uplink phase, $N_3$ can reliably obtain $W_1 +_q W_2$ with a rate approaching $\frac{1}{2}\log_2(1+\min(P_1,P_2))$.

***Proposition 4:*** In high SNR region, $N_3$ can obtain $W_1 +_q W_2$ at a rate approaching $\frac{1}{2}\log_2(1+\min(P_1,P_2))$ with the PNC scheme depicted in Fig. 3 if the modulo-$q$ LDPC codes can approach the Gaussian channel capacity in high SNR region.

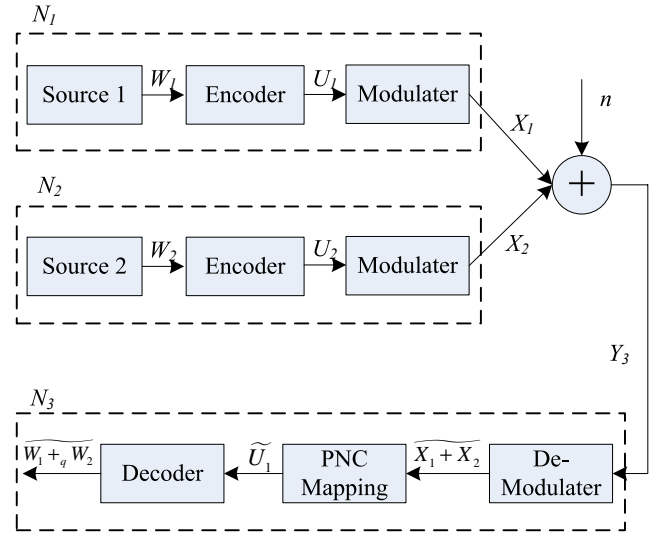

Figure 3. Transmission and reception diagram of PNC scheme

*System Description:* Before the proof, let us try to understand the proposed system first. With reference to Fig. 3, let us assume $P_1 \leq P_2$ without loss of generality. The transmission strategy is as follows. We first reduce the transmission power of node $N_2$ to $P_1$ so that both end nodes transmit with the same power. We use the same block channel code, which is linear with the operation of modulo-$q$ addition, at the two end nodes. In other words, we have

$$\Gamma_1 = \Gamma_2 = \Gamma$$
$$U_1 +_q U_2 = \Gamma(W_1) +_q \Gamma(W_2) = \Gamma(W_1 +_q W_2) \quad (11)$$

Therefore, the length of $U_1$ equals the length of $U_2$. The modulo-$q$ LDPC code, first proposed by G. Gallager in [14], is an example of such a linear coding scheme. We also assume that the $q$-ary PAM modulation (QAM modulation is also valid in the PNC system) is used at both end nodes, i.e.,

$$X_i = M(U_i) = 2U_i - (q-1) \quad (12)$$

The two end nodes cooperate to transmit so that the two transmitted signals reach node $N_3$ in synchrony. Therefore, we can denote the $k$-th received symbol on baseband as

$$y_3[k] = x_1[k] + x_2[k] + n \quad (13)$$

After receiving $Y_3$, $N_3$ can obtain an estimation of $X_1 + X_2$, with either hard detection or soft detection, denoted by $\widetilde{X_1 + X_2}$. Although optimal demodulation thresholds could be used to obtain $\widetilde{X_1 + X_2}$ (see [8]), we will show that using the middle value of two adjacent constellation points as the estimation threshold is good enough to prove *Proposition 4*. After that, $N_3$ will map $\widetilde{X_1 + X_2}$ to the estimation of $U_1 +_q U_2$, denoted by $\widetilde{U_1 +_q U_2}$ with a PNC mapping, as follows

$$\widetilde{U_1 +_q U_2} = (\widetilde{X_1 + X_2})/2 - 1 \bmod q \quad (14)$$

Due to the linearity of $\Gamma$, $U_1 +_q U_2$ is the codeword of $W_1 +_q W_2$ as shown in (11). Therefore, we can decode $\widetilde{U_1 +_q U_2}$ with the decoding process corresponding to $\Gamma$ to obtain an estimation of the target information, denoted by $\widetilde{W_1 +_q W_2}$. Note that $W_1$ and $W_2$ are not decoded explicitly in the process to obtain $\widetilde{W_1 +_q W_2}$.

A critical assumption in this proposition is that the modulo-$q$ LDPC code can approach the Gaussian channel capacity in high SNR region. This assumption is supported by the following observations. First, [15] proved that modulo-$q$ quantized coset (MQC) LDPC codes can achieve the Shannon capacity of any discrete memoryless channel and the simulation there showed that MQC-LDPC codes can approach the capacity of AWGN channel within a gap of 0.9 dB. Second, for modulo-$q$ LDPC code, the shaping loss (compared to the MQC LDPC) due to the lack of quantization is at most 1.53dB as shown in [16]. Such a fixed SNR loss can be ignored in high SNR region. We now prove *Proposition 4* with the introduced PNC transmission scheme.

**Proof:** First consider the point-to-point transmission in Fig. 4. Suppose the transmission power of the source is also $P_1$, and other assumptions are same as in our system model. As shown in [17, Ch5.2], the symbol error rate (SER) before channel decoding is

$$P_q = \Pr(u \neq \tilde{u}) = \Pr(x \neq \tilde{x}) = \frac{q-1}{q}\Pr(|n| \geq d/2) \quad (15)$$

where $d$ is the distance between adjacent constellation points and $\Pr(|n| \geq d/2)$ is the probability that the Gaussian noise will cause a symbol detection error. Since we assume that the coset modulo-$q$ LDPC code with $q$-ary PAM modulation can approach the Gaussian channel capacity in high SNR region, the SER after decoding, i.e., $\Pr(w \neq \tilde{w})$, will go to zero only if the information rate, determined by the LDPC code rate, does not exceed the Gaussian channel capacity $\frac{1}{2}\log_2(1+P_1)$ in high SNR region.

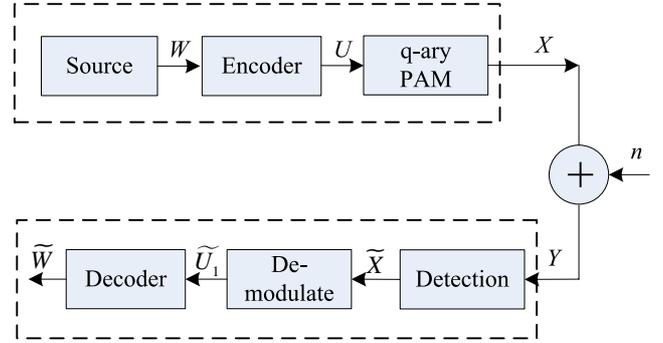

Figure 4. Point-to-point transmission

We now apply the identical coset-LDPC code at $N_1$ and $N_2$ in our system of Fig. 3. Note that the coset vector in the point-to-point transmission can be achieved by the two cooperatively designed coset vectors at $N_1$ and $N_2$. And we also set the code rate to be the same as that in the point-to-point transmission in Fig. 4. The SER of PNC scheme before decoding is

$$\begin{aligned} P_{q+} &= \Pr((u_1[k] +_q u_2[k]) \neq \widetilde{(u_1[k] +_q u_2[k])}) \\ &< \Pr((x_1[k] +_q x_2[k]) \neq \widetilde{(x_1[k] +_q x_2[k])}) \\ &= \frac{q^2-1}{q^2}\Pr(|n| \geq d/2) \end{aligned} \quad (16)$$

where the last line is derived in the paragraph after this proof. As SNR and $q$ go to infinity, we can see that $P_{q+}$ in (16) will go to $P_q$. Since the same channel encoding and decoding scheme is used, the PNC SER after channel decoding will also go to zero as in the point-to-point system. Because the code rate is same as that of the point-to-point transmission, the information rate of $\widetilde{W_1 +_q W_2}$, which is equal to that of $W_1$ and $W_2$, also approaches $\frac{1}{2}\log_2(1+P_1)$.

*SER of two superimposed PAM signal*

For a superposition of two synchronized $q$-ary PAM signals, the new constellation still looks like that of PAM,

where the probability of the two end points is $1/q^2$, and the probability of all other constellation points is $1-2/q^2$. Since the two PAM signals have the same transmitting power, the distance between two adjacent constellation points of the superimposed signal is the same as that of the single PAM signal case, denoted by $d$. The SER of the left and right end points are

$$\Pr_1 = \Pr(n < -d/2)$$
$$\Pr_2 = \Pr(n > d/2) \qquad (17)$$

where $n$ is the Gaussian noise. The SER of other points is

$$\Pr_3 = \Pr(|n| > d/2) \qquad (18)$$

The overall SER is

$$P = \frac{1}{q^2}\Pr_1 + \frac{1}{q^2}\Pr_2 + \frac{q^2-1}{q^2}\Pr_3$$
$$= \frac{q^2-1}{q^2}\Pr(|n| \geq d/2) \qquad (19)$$

Finally, combining the PNC multiple access scheme and network-coding-like broadcasting scheme, the upper bound of the TWRC capacity is approachable in high SNR region.

## VI. CONCLUSION

In this paper, we have investigated the capacity of three-node TWRC, in which two end nodes exchange information via a relay node. An upper bound of the capacity has been given, and the tightness of the bound has been proved in the low SNR and high SNR regions.

We conjecture that the high SNR limitation in *Proposition 4* is not necessary for the following reasons. First, we may carefully design the constellation mappings at the two end nodes so that the superimposed signal at the relay can achieve the shaping gain. Second, with reference to the upper bound proof in the last section, we could fix SNR and increase $q$ in (15) and (16). This has the effect of increasing the factor $\Pr(|n| \geq d/2)$ in (15) and (16) similarly. Meanwhile, both the factors $\frac{q-1}{q}$ and $\frac{q^2-1}{q^2}$ approach one. Third, simulation results in [18] suggest that by increasing $q$ in PNC, the achievable rate may be able to approach channel capacity.

This paper has assumed a set-up in which a direct link between the two end nodes is unavailable. We conjecture that this assumption is not necessary for the results we have obtained as long as links are half-duplex (i.e., a node cannot receive and transmit at the same time), and the capacity of the direct link between the two end nodes is no better than half of either of the relay links between the end nodes and the relay node.

The PNC uplink transmission scheme in Fig. 3 can be viewed as a specific scheme for "computation over multiple access channels" first studied in [19]. For the computation over multiple-access channels, the receiving node is interested in a function of the inputs, instead of each individual input. Theoretical results on the capacities of such multiple-access channels for certain functions of inputs have been obtained in [19]. The case of the function as modulo-$q$ addition of the inputs, such as in our PNC scheme in Fig. 3, is first addressed in our paper here.